\documentclass[journal=jctcce,manuscript=article]{achemso}
\setkeys{acs}{usetitle=true}

\usepackage[version=3]{mhchem} % Formula subscripts using \ce{}

\usepackage{amsmath,amssymb}
\usepackage{graphicx,color}
\usepackage{amsfonts}
\usepackage[outdir=./]{epstopdf}
\usepackage[linesnumbered,ruled]{algorithm2e}
\usepackage{comment}
\usepackage{placeins}
\usepackage{fixltx2e}
\usepackage{epstopdf}
\usepackage{subfig} 
\allowdisplaybreaks

\usepackage[version=3]{mhchem} % Formula subscripts using \ce{}

\global\long\def\P{\mathbb{P}}

\global\long\def\Tr{\mathrm{Tr}}

\renewcommand{\Im}{\mathrm{Im}}

\newcommand{\norm}[1]{\lVert#1\rVert}

\newcommand{\ie}{\textit{i.e.}~}
\newcommand{\eg}{\textit{e.g.}~}

\newcommand{\I}{\imath}

\newcommand{\angstrom}{\mbox{\normalfont\AA}~}

\title[Fast RT-TDDFT]{Fast real-time time-dependent \\
density functional theory calculations \\
with the parallel transport gauge}% Force line breaks with

\author{Weile Jia}
\affiliation{Department of Mathematics, University of California,
Berkeley, California 94720, United States}

\author{Dong An}
\affiliation{Department of Mathematics, University of California,
Berkeley, California 94720, United States}

\author{Lin-Wang Wang}
\affiliation{Materials Science 
Division, Lawrence Berkeley National Laboratory, Berkeley, California 94720, United States}

\author{Lin Lin}
\email{linlin@math.berkeley.edu}
\affiliation{Department of Mathematics, University of California,
Berkeley, California 94720, United States}
\affiliation{Computational Research Division, Lawrence Berkeley
National Laboratory, Berkeley, California 94720, United States}

\begin{document}

\begin{abstract}
Real-time time-dependent density functional theory (RT-TDDFT)
is known to be hindered by the very small time step (attosecond
or smaller) needed in the numerical simulation due to the fast oscillation 
of electron wavefunctions, which significantly limits its
range of applicability for the study of ultrafast dynamics. In this
paper, we demonstrate that such oscillation can be considerably reduced 
by optimizing the gauge choice using the parallel transport formalism. 
RT-TDDFT calculations can thus be significantly accelerated using a combination of 
the parallel transport gauge and implicit integrators, and the resulting
scheme can be used to accelerate any electronic structure software that
uses a Schr\"odinger representation.
Using absorption spectrum, ultrashort
laser pulse, and Ehrenfest dynamics calculations for example,
we show that the new method can utilize a time step that is on the
order of $10\sim 100$ attoseconds in a planewave basis set, and is no
less than $5\sim 10$ times faster when compared to the standard
explicit 4th order Runge-Kutta time integrator. 

% Real-time time-dependent density functional theory (RT-TDDFT)
% is known to be hindered by the very small time step (attosecond
% or smaller) needed in the numerical simulation, which significantly limits its
% range of applicability for the study of ultrafast dynamics. In this
% paper, we demonstrate that RT-TDDFT calculations can be significantly
% accelerated using a combination of two techniques: parallel transport
% gauge and implicit time integrator, while neither technique would be
% sufficiently effective on its own. Using absorption spectrum, ultrashort
% laser pulse, and Ehrenfest dynamics calculations for example,
% we show that the new method can utilize a time step that is on the
% order of $10\sim 100$ attoseconds  in a planewave basis set, and is no
% less than $5\sim 10$ times faster when compared to standard explicit
% time integrators.

%\begin{description}
%\end{description}
\end{abstract}

\section{Introduction}

Recent developments of ultrafast laser techniques have enabled a large
number of excited state phenomena to be observed in real time.
One of the most widely used techniques for studying ultrafast
properties is the real-time time-dependent density functional theory
(RT-TDDFT)~\cite{RungeGross1984,OnidaReiningRubio2002}, which has achieved successes in a number
of fields including \eg nonlinear optical
response~\cite{TakimotoVilaRehr2007} and the collision
of an ion with a substrate~\cite{KrasheninnikovMiyamotoTomanek2007}.  
Nonetheless, the range of applicability of RT-TDDFT is often hindered by
the very small time step needed to propagate the Schr\"odinger equation.
Many numerical propagators used in practice are explicit time
integrators~\cite{CastroMarquesRubio2004,SchleifeDraegerKanaiEtAl2012,GomezMarquesRubioEtAl2018},
which require a small time step size satisfying $\Delta t \lesssim
\norm{H}^{-1}$ due to the stability restriction. 
For $H$ discretized under a flexible basis set such as
planewaves, the required time step is often less than $1$ attosecond (as). On the other hand,
ultrafast properties often need to be observed on the order of $10\sim
10^3$ femtoseconds (fs). This requires $10^{4}\sim 10^{6}$ time steps to be performed and is often prohibitively expensive. 
Given the recent emphasis on ultrafast physics, this is thus an urgent 
problem to be solved. 
%While implicit time integrators are known to be
%numerically stable with even large time step for stiff problems, it is
%not a very effective strategy to use such integrators directly for
%RT-TDDFT either. This is because each element of the Kohn-Sham orbital is
%often oscillating on the attosecond or sub-attosecond timescale in the
%complex plane,thus a time step on the scale of attosecond is still required to 
%resolve the fast oscillation and to obtain an accurate numerical approximation. 

However, physical observables such as the electron density 
are squared quantities of the wavefunctions, and often
oscillate much slower. In this paper, we find
that such gap is largely due to the non-optimal gauge choice of the
Schr\"odinger dynamics,  which is irrelevant to the computation of
physical observables. We propose that the optimal gauge choice is given
by a parallel transport formulation. Compared to the Schr\"odinger
representation, the orbitals with the parallel transport gauge can often
be ``flattened'' into an approximate straight line over a much longer
time interval.   When combined with implicit time integrators to
propagate the parallel transport dynamics, it is possible to
significantly increase the time step size without sacrificing 
accuracy.  The parallel transport formulation only introduces one extra
term to the Schr\"odinger equation, and thus can be easily applied to any
electronic structure software packages for RT-TDDFT calculations, 
which is unlike to other methods where approximations and significant 
rewriting are needed~\cite{JahnkeLubich2003,WangLiWang2015}.

\section{Theory}

In order to derive the parallel transport gauge, let us first consider
the RT-TDDFT equations
\begin{equation} \I \partial_{t} \psi_{i}(t) = H(t,P(t)) \psi_{i}, \quad
  i=1,\ldots,N_{e}.
  \label{eqn:tddft}
\end{equation}
Here $\Psi(t)=[\psi_{1},\ldots,\psi_{N_{e}}]$ are the electron orbitals,
and the Hamiltonian can depend explicitly on $t$ and nonlinearly on the
density matrix $P(t) = \Psi(t)\Psi^{*}(t)$ or the electron density 
$\rho(t) = \sum_{i=1}^{N_e}|\psi_i(t)|^2$. 
% For local and semilocal exchange-correlation
% functionals~\cite{PerdewZunger1981,PerdewBurkeErnzerhof1996}, the dependence on
% $P$ can be further simplified to that on the electron density
% $\rho(t)$ only. Hybrid exchange-correlation
% functionals~\cite{Becke1993,HeydScuseriaErnzerhof2003} involve a
% fraction of the Fock exchange operator and thus depend on the
% entire density matrix $P(t)$.
Eq.~\eqref{eqn:tddft} can be equivalently written using a set of
transformed orbitals $\Phi(t)=\Psi(t) U(t)$, where the gauge matrix $U(t)$ is a
unitary matrix of size $N_{e}$.
An important property of the density matrix is that it is
gauge-invariant: $P(t)=\Psi(t)\Psi^{*}(t)=\Phi(t)\Phi^{*}(t)$, and always
satisfies the von Neumann equation (or quantum Liouville equation)
\begin{equation}
%  \I \partial_{t} P(t) = [H(t,P),P(t)] = H(t,P(t))P(t) - P(t)H(t,P(t)).
  \I \partial_{t} P = [H,P] = HP - PH.
  \label{eqn:vonNeumann}
\end{equation}

Our goal is to optimize the gauge matrix, so that 
the transformed orbitals $\Phi(t)$ vary \textit{as slowly as
possible}, without altering the density matrix. 
This results in the following variational problem
\begin{equation}\label{eqn:minproblem}
  \min_{U(t)} \quad \norm{\dot{\Phi}}^2_{F},
  \ \text{s.t.} \ \Phi(t) = \Psi(t)U(t), U^{*}(t)U(t)=I_{N_{e}}.
%\begin{split}
%  \min_{U(t)} \quad &\norm{\dot{\Phi}}^2_{F} \\
%  \text{s.t.} \quad &\Phi(t) = \Psi(t)U(t), \quad U^{*}(t)U(t)=I.
%\end{split}
\end{equation}
Here $\norm{\dot{\Phi}}^2_{F}:=\mathrm{Tr}[\dot{\Phi}^{*}\dot{\Phi}]$
measures the Frobenius norm of the time derivative of the transformed
orbitals.  The minimizer of~\eqref{eqn:minproblem}, in terms of $\Phi$, 
satisfies~\footnote{See Appendix A for its derivation, also for an explanation on the name ``parallel transport''}
\begin{equation}
  P\dot{\Phi}=0 {.}
  \label{eqn:PTcondition}
\end{equation}
Eq.~\eqref{eqn:PTcondition} implicitly defines a gauge
choice for each $U(t)$, and this gauge is called the \emph{parallel transport
gauge}.  The governing equation of each transformed orbital $\varphi_{i}$ can be concisely written down as~\footnote{See Appendix A for its derivation, also for an explanation on the name ``parallel transport''}
\begin{equation}
  \I \partial_t \varphi_i = H\varphi_i - 
  \sum_{j=1}^{N_e}\varphi_j\left<\varphi_j|H|\varphi_i\right>, 
  \quad i = 1, \cdots, N_e,
  \label{eqn:pt-ket}
\end{equation}
or more concisely in the matrix form
\begin{equation}
  \I \partial_t \Phi = H\Phi -
  \Phi(\Phi^{*}H\Phi), \quad P(t) = \Phi(t)\Phi^{*}(t).
  \label{eqn:pt}
\end{equation}
%or equivalently in a band index explicit way with bra-ket notation 
%\DA{Here Lin-Wang suggests to use the band index explicit way with bra-ket notation 
%instead of math writing style. I remain both of them, because the notations of the 
%numerical schemes can be very complicated 
%with band index explicit way and bra-ket notation. }
% \begin{equation}
%   \I \partial_t \Phi = H\Phi -
%   \Phi(\Phi^{*}H\Phi), \quad P(t) = \Phi(t)\Phi^{*}(t).
%   \label{eqn:pt}
% \end{equation}
%The parallel transport evolution of Eq.~\eqref{eqn:pt} only
%introduces one additional term $\Phi(\Phi^{*}H\Phi)$ 
%compared to~\eqref{eqn:tddft}. 
The right hand side of Eq.~\eqref{eqn:pt} is analogous to the residual vectors of an eigenvalue
problem in the time-independent setup.  
Hence $\Phi(t)$ follows the dynamics driven by residual vectors and is
expected to vary slower than $\Psi(t)$.

\begin{figure}[h]
  \begin{center}
    \subfloat[Real parts of $\psi(x_{0},t)$ and $\varphi(x_{0},t)$ at
    $x_0 = 25.0$]{\includegraphics[width=0.45\textwidth]{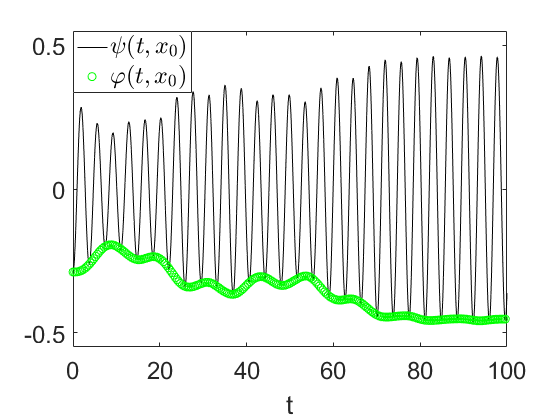}}
    \subfloat[Dipole moment]{\includegraphics[width=0.45\textwidth]{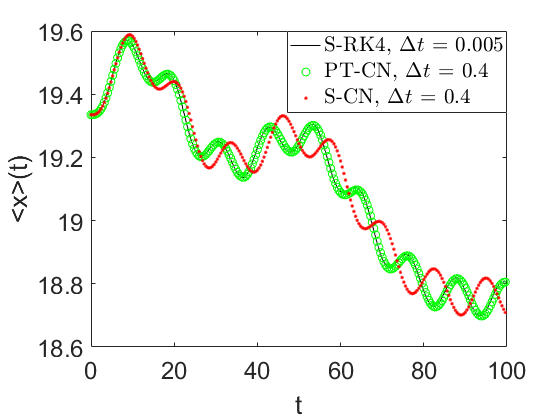}}
  \end{center}
  \caption{Comparison of the Schr\"odinger and parallel transport dynamics
  for the 1D example.}
%  the real parts of $\psi(x_{0},t)$ and
%  the parallel transport variable $\varphi(x_{0},t)$ for $x_0 = 25.0$.
%  (b) Center of wavefunction obtained respectively 
%  by applying S-RK4 to Schr\"odinger equation, GL2 to PT dynamics 
%  and GL2 to Schr\"odinger equation.
  \label{fig:schtoy}
\end{figure}

The advantage of the parallel transport gauge is most clear 
in the \textit{near adiabatic regime}, when the
right hand side of~\eqref{eqn:pt} approximately vanishes. 
% Hence the oscillation of $\Phi(t)$ can be significantly slower than that of
% $\Psi(t)$. 
Fig.~\ref{fig:schtoy} (a) demonstrates a simple example with
one electron, and a time dependent Hamiltonian 
$H(t)=-\frac12 \partial^2_x + V(x,t)$ in one dimension. The initial state is 
the ground state of $H(0)$. 
Here the time-dependent potential is chosen to be
$V(x,t) = -2\exp(-0.1(x-R(t))^2) -2\exp(-0.1(x-12.5)^2)$, 
which is a double well potential with 
one fixed center at $12.5$ and 
one moving center  $R(t) = 25 + 1.5\exp(-0.0025(t-10)^2) + \exp(-0.0025(t-50)^2)$. 
Fig.~\ref{fig:schtoy} (a) shows that while $\psi(t)$ oscillates rapidly,
the oscillation of the parallel transport orbital $\varphi(t)$ is significantly slower, and
can thus be approximated by a straight line over a much larger
interval.  We remark that efficient numerical methods based on the
construction of instantaneous adiabatic states have also been recently
developed for the near adiabatic regime~\cite{JahnkeLubich2003,WangLiWang2015}.  
The advantage of the
parallel transport dynamics is that it only operates on $N_{e}$ orbitals
as in the original Schr\"odinger dynamics.  Even outside the near adiabatic
regime, Eq.~\eqref{eqn:pt} always yields the slowest possible dynamics
due to the \textit{variational principle} in Eq.~\eqref{eqn:minproblem}.

In order to propagate the parallel transport dynamics numerically, 
all the RT-TDDFT propagation methods can be used 
since Eq.~\eqref{eqn:pt} only differs from Eq.~\eqref{eqn:tddft} 
in one extra term $\Phi(\Phi^{*}H\Phi)$. 
% In order to propagate the parallel transport dynamics numerically with a
% relatively large time step, we should use implicit time integrators
% especially when $\norm{H}$ is relatively large. 
%Standard implicit time integrators, such as the Crank-Nicolson scheme
%(CN), can be numerically stable even if
%the time step is very large. However, when applied to RT-TDDFT these
%schemes suffer from significant loss of accuracy. This is because such
%integrators aims at approximating the solution within a time interval
%$(t,t+\Delta t)$ using a low order polynomial. For instance, CN
%uses a trapezoidal rule for time discretization, which approximates the
%solution by a straight line connecting the two end points.  This
%approximation is doomed to fail when $\Psi(t)$ is fast oscillating as in
%Fig.~\ref{fig:}. On the other hand, the effectiveness of such linear
%approximation is much improved when applied to $\Phi(t)$ as in
%Fig.~\ref{fig:}.  This allows us to use a significantly larger time step
%without sacrificing the accuracy requirement. 
As an example, the parallel transport Crank-Nicolson scheme (PT-CN)
gives rise to the following scheme
%In order to solve Eq.~\eqref{eqn:pt} numerically, we should not use
%explicit integrators. This is because the energy spectrum of the
%right hand side is still close to $\norm{H}$, which limits the step size
%to the order of $10^{-3}$ fs. However, since $\Phi(t)$ varies much
%smoother than $\Psi(t)$, we can utilize implicit integrators to overcome
%the problem of the small step size. For instance, the trapezoidal rule
%for discretizing Eq.~\eqref{eqn:pt} reads
%\begin{widetext}
%\begin{equation}\label{eqn:ptcn}
%    \Phi_{n+1} + \I \frac{\Delta t}{2} \left\{
%    H_{n+1} 
%    \Phi_{n+1} - \Phi_{n+1}\left(\Phi_{n+1}^{*}
%    H_{n+1} \Phi_{n+1}\right)\right\} 
%    =  \Phi_{n} -\I \frac{\Delta t}{2} \left\{
%    H_{n}
%    \Phi_{n} - \Phi_{n}\left(\Phi_{n}^{*}
%    H_{n} \Phi_{n}\right)\right\},
%\end{equation}
%\end{widetext}
\begin{align}\label{eqn:ptcn}
    &\Phi_{n+1} + \I \frac{\Delta t}{2} \left\{
    H_{n+1} 
    \Phi_{n+1} - \Phi_{n+1}\left(\Phi_{n+1}^{*}
    H_{n+1} \Phi_{n+1}\right)\right\} \nonumber \\
    = & \Phi_{n} -\I \frac{\Delta t}{2} \left\{
    H_{n}
    \Phi_{n} - \Phi_{n}\left(\Phi_{n}^{*}
    H_{n} \Phi_{n}\right)\right\}.
\end{align}
Here $H_{n}=H(t_{n},P_{n})$ is the Hamiltonian at the time step $t_{n}$, and
$t_{n+1}=t_{n}+\Delta t$. Other time integrators
can be straightforwardly generalized to the parallel transport dynamics as well
(see Appendix B).
In Eq.~\eqref{eqn:ptcn}, the solution $\Phi_{n+1}$ needs to be solved
self-consistently. This is a set of nonlinear equations with respect to the unknowns
$\Phi_{n+1}$, and can be efficiently solved by \eg the preconditioned
Anderson mixing scheme~\cite{Anderson1965}.  The propagation of $\Phi(t)$ can also
be naturally combined with the motion of nuclei discretized \eg by the
Verlet scheme for the simulation of Ehrenfest
dynamics~\cite{Lubich2008book}.

\begin{table}
  \centering
  \begin{tabular}{c|c|c}
  \hline Method & $\Delta t$ & Error \\\hline
  S-RK4  & 0.01   & $3.76\times10^{-7}$   \\
  PT-RK4 & 0.01   & $4.36\times10^{-10}$  \\
  S-RK4  & 0.005  & $2.35\times10^{-8}$   \\
  PT-RK4 & 0.005  & $2.73\times10^{-11}$  \\
  \hline
  \end{tabular}
  \caption{Error at $T = 100$ for 
  the 1D example. 
  }\label{tab:OrbitalErrors}
\end{table}

Since the parallel transport dynamics yields the optimal gauge choice, 
it can be used to improve the accuracy and
efficiency of any propagator currently applied to the Schr\"odinger dynamics.
For example, let us first consider again the one dimensional Schr\"odinger equation.
Table~\ref{tab:OrbitalErrors} compares 
the explicit 4th order Runge-Kutta scheme for the Schr\"odinger dynamics (S-RK4) 
and the parallel transport dynamics (PT-RK4), 
which indicates that the error of the latter is considerably smaller. 
When combined with implicit integrators, we can further significantly increase the
time step size. 
% In order to demonstrate the effectiveness of the combined strategy of
% parallel transport gauge and implicit time integrator, 
We compare the PT-CN 
scheme with the standard CN scheme for the Schr\"odinger dynamics (S-CN), and use S-RK4 
with a small time step $\Delta t=0.005$ as the benchmark. The
accuracy is measured by the dipole moment $\langle x\rangle(t)
= \mathrm{Tr}[x P(t)]$ along the trajectory. 
Fig.~\ref{fig:schtoy} (b) indicates that PT-CN can use a much larger
time step than S-RK4 without losing accuracy. On the other hand, while S-CN
can still be numerically stable with the same step size, it becomes
significantly less accurate after the first peak around $t=10$.

\section{Numerical results}

We demonstrate the performance of the PT-CN scheme for RT-TDDFT
calculations for three real systems representing three prototypical 
usages of RT-TDDFT. 
Our method is implemented in PWDFT code, which uses the
planewave basis set and is a 
self-contained module in the massively parallel DGDFT (Discontinuous
Galerkin Density Functional Theory) software
package~\cite{LinLuYingE2012,HuLinYang2015a}.
We use the Perdew-Burke-Ernzerhof (PBE) exchange correlation
functional~\cite{PerdewBurkeErnzerhof1996}, and the Optimized
Norm-Conserving Vanderbilt (ONCV)
pseudopotentials~\cite{Hamann2013,SchlipfGygi2015}. 

\begin{figure}[h]
  \begin{center}
    \includegraphics[width=0.50\textwidth]{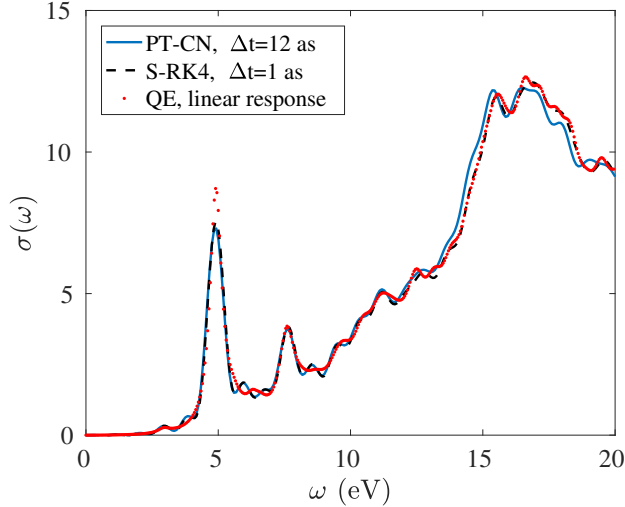}
  \end{center}
  \caption{Absorption spectrum for anthracene.}
  \label{fig:absorp_anthracene}
\end{figure}

The first example is the computation of the absorption spectrum of
an anthracene molecule.
We set the time step size of PT-CN to be $12$ attoseconds (as), 
and that of S-RK4 to be 1 as 
(it becomes unstable when the step size is larger). 
Fig.~\ref{fig:absorp_anthracene} compares the absorption spectrum
obtained from PT-CN and S-RK4 with PWDFT. This result is benchmarked
against the linear response time-dependent density functional theory
(LR-TDDFT) calculation using the turboTDDFT
module~\cite{MalciogluGebauerRoccaEtAl2011} from the
Quantum ESPRESSO software package~\cite{GiannozziBaroniBoniniEtAl2009}, which performs $3000$ Lanczos
steps along each perturbation direction to evaluate the polarization
tensor. A Lorentzian smearing of $0.27$ eV is applied to all
calculations. We find that the absorption spectrum calculations from the
three methods agree very well. The spectrum obtained from PT-CN and that
from S-RK4
are nearly indistinguishable below $10$ eV, and becomes slightly
different above $15$ eV.  Note that the $\delta$-pulse simultaneously
excites all eigenstates from the entire spectrum, and $\omega=15$ eV
already amounts to the time scale of $40$ as, which is approaching the
step size of the PT-CN method. 
Since the computational cost of RT-TDDFT calculations is mainly dominated
by the cost of applying the Hamiltonian operator to orbitals, 
we measure the numerical efficiency using the number of such
matrix-vector multiplications per orbital.
The PT-CN method requires on average
$4.9$ matrix-vector multiplications for each orbital. This is
comparable to the S-RK4 method which requires $4$ matrix-vector
multiplications per time step. Hence for this example, the PT-CN method
is around $10$ times faster than the S-RK4 method.

%test our algorithm, we used two systems respectively for absorption spectrum and 
%ion collision test. The Perdew, Burke, Ernzerhof \ref{PBE} gerneralized 
%gradient appromixation and plane wave basis set are used in the tests. 
%The SG15 Optimized Norm-Conserving Vanderbilt (ONCV) pseudopotentials 
%is used to converge the ground state in our PWDFT code \ref{PWDFT}. 
%We always use a $\Gamma$ point calculation in the following tests. 

%\LL{Organize the benzene example, and add the example of ion collision}

% change from benzene to antho
%\begin{figure}
%\begin{center}
%\includegraphics[width=0.4\textwidth]{vmdbenzene.png}
%\end{center}
%\caption{Model of benzene. The external electric field is along the x axis. }
%\label{fig:benzenemodel}
%\end{figure}

\begin{figure}[h]
  \begin{center}
    \subfloat[]{\includegraphics[width=0.45\textwidth]{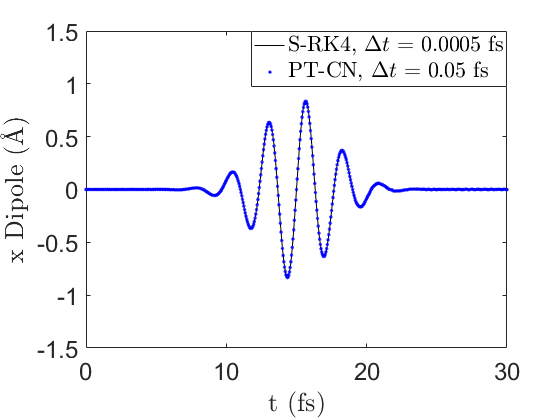}}
    \subfloat[]{\includegraphics[width=0.45\textwidth]{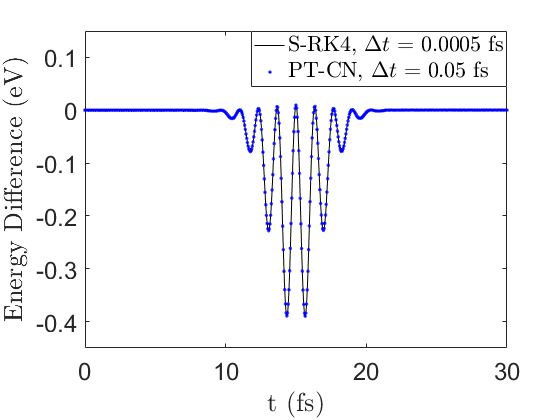}}\\
    \subfloat[]{\includegraphics[width=0.45\textwidth]{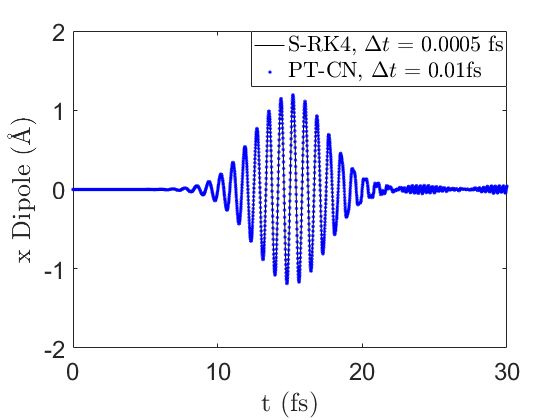}}
    \subfloat[]{\includegraphics[width=0.45\textwidth]{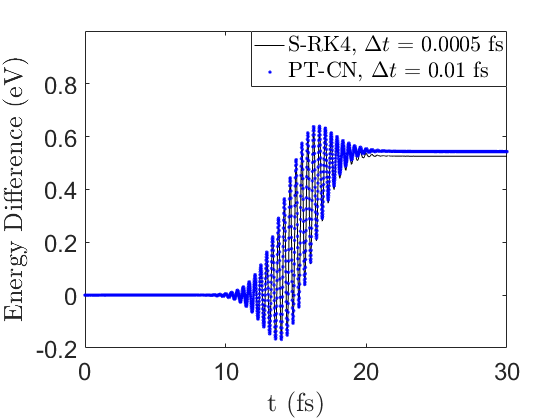}}
  \end{center}
  \caption{Electronic dynamics of benzene driven by
  laser with 
  $\lambda = 800$ nm in (a)(b), and 250 nm in (c)(d). }
  \label{fig:benzeneresults}
\end{figure}

%The second system is a 12 atom benzene system with external electric field 
%in the X direction, as shown in Fig.~\ref{fig:benzenemodel}.
%We have used 15 states in the calculation, each state 
%is occupied with 2 electrons. Note that 15 MPIs is used to parallelize the
%15 states. We tested two lasers, the first one with a wave length of 800nm
%and the second with a 250nm wave length (see Fig.~\ref{fig:benzeneVext}). 
%Time step is 0.05fs for the first 
%test and 0.01fs for the second test. Since benzene has a wide band gap
%(5.23 eV). We can only observe the excited states with high frequency laser. 
%We compared this system with both our S-RK4 method implemented in PWDFT with a
%time step of 0.5 as, as shown in Fig.~\ref{fig:benzeneresults}. 
%For the ground state computation, the difference is extremely small, and 
%PT-CN reduces the total number of matrix-vector multiplications by a 
%factor of 31.7. 
%For the second case, PT-CN still matches very well during the 
%physical relevant time, slightly overestimates the total energy 
%by 0.0175 eV at the end of the simulation, 
%and gives a speedup of 9.6. 

The second system is a benzene molecule driven by an ultrashort laser
pulse.  We apply two
lasers with its wavelength being 800 nm and 250 nm, respectively.    
We measure the accuracy using the dipole moment along the
x direction, as well as the energy difference $E(t)-E(0)$ along the
trajectory, as shown in Fig.~\ref{fig:benzeneresults}. 
For the case of the 800 nm laser, 
Fig.~\ref{fig:benzeneresults} confirms that the results of PT-CN with a
time step size of $50$ as fully
match those obtained from the S-RK4 method with a time step that is 100 times smaller. 
Again S-RK4 becomes unstable when the time step size is larger than $1$ as. 
After 25.0 fs, the increase of the total energy
for S-RK4 and PT-CN is $2.00\times 10^{-4}$ eV and $2.44\times 10^{-4}$
eV, respectively, indicating negligible energy absorption 
due to the band gap of the system. 
During the time interval for which the
laser is active (from 5.5 fs to 24.5 fs), 
the average number of matrix-vector multiplications per orbital
in each PT-CN time step is 12.6, 
and the total number of matrix-vector multiplications per orbital is 4798. 
The total number of matrix-vector multiplications per orbital for S-RK4 within the
same time interval is 152000, and the speedup of PT-CN over RK4 is $31.7$.

The 250 nm laser has a higher photon energy above the benzene band gap, 
and thus results in significant energy absorption. 
In this case, even physical observables such as dipole moments become fast
oscillating, and PT-CN needs to adopt a smaller 
step size $10$ as 
% \LL{5 as or 10 as?}
% \DA{Here in Figure 3 
% I use the example with 10 as, because I see in the abstract 
% that the target here is to achieve time steps on the order of 10-100 as, 
% and I think the accuracy of the numerical results obtained with dt = 10 as 
% is still tolerable. } \LL{OK. please make sure all results to be
% consistent here},
and still yields very good approximation to the
electron dynamics compared to S-RK4.
The increase of the total energy after 25.0 fs 
for S-RK4 and PT-CN is $0.526$ eV and $0.544$ eV, respectively. 
The average number of matrix-vector multiplications  per orbital
in each PT-CN time step is 8.3 due to the reduced step size, 
and the total number of matrix-vector multiplications per orbital is 15817.  
Therefore in this case PT-CN achieves 9.6 times speedup over S-RK4.

As the last example, we use the RT-TDDFT based Ehrenfest dynamics to study 
the process of a chlorine ion (Cl$^{-}$) colliding to a graphene nanoflake 
consisting of 112 atoms. This models
the ion implantation procedure for doping a substrate. 
At the beginning of the simulation, the Cl$^{-}$ is placed at 6 \angstrom away from 
the graphene and is given an initial velocity perpendicular to the
plane of the graphene pointing towards the center of one hexagonal ring
formed by the carbon atoms. 
%The initial temperature is set 
%to 0K to avoid random thermomovement\LL{why we need an initial
%temperature?} \WL{Cause the substrate velocity is set to be zero, which corresponds to
%a 0K temperature. }
The initial velocity of Cl$^{-}$ is set to be $0.5$, $0.75$, $1.0$, $1.5$,  $2.0$,
$4.0$ and $8.0$ Bohr/fs, and the kinetic energy carried by
the ion ranges from $128$ eV to $32926$ eV, respectively. 
The simulation is terminated before the ion reaches the boundary of
the supercell. For instance, we set $T=10$ fs when the velocity is $2.0$
Bohr/fs. In such case, the time step size for PT-CN and S-RK4 is set to be $50$
as and $0.5$ as, respectively.  Each PT-CN step requires on average
$28$ matrix-vector multiplication operations per orbital, and the overall speedup of
PT-CN over S-RK4 is $14.2$.

%We only simulate
%enough time for the projectile to pass through the substrate, i.g., 10fs for 
%2.0 Bohr/fs case, although longer simulation time is possible. 

%2 dimensional 112 atom monolayer of graphene sheet
%the tested the TDDFT method on a chlorine atom (Cl- ion) colliding to a 
%2 dimensional 112 atom monolayer of graphene sheet. 
%Graphene is important 2D
%material with potential usage in many fields, such as transistors and renewable 
%energy applications\ref{}. In the electronic industry, one standard way of 
%doping the system is ion implantation, where an ion beam directly collide with
%substrate. Ion and graphene collision has been studied with TDDFT method before \ref{}.

%PTCN method is much faster compared with direct S-RK4 method. For example, 
%in the case of $2.0$ Bohr/fs test,PTCN time step is chosen to be 0.05fs, 
%while the S-RK4 direct method time step is 0.5as, so in this particular 
%testing case our method can increase the time step by a factor of 100. 
%Such big time step gap remains for other tests. Please note that our 
%algorithm is still stable with bigger time step, i.g., 0.1fs, 
%however, the DIIS steps to converge will increase by a factor 
%of 2.2x for this system. Thus 0.05fs is a good choice.  

%It costs less than 5 hours using 228 cores on Edison supercomputer 
%at National Energy Research Scientific Computing Center(NERSC) to finish 
%the 2.0 bohr/fs case. One TDDFT step(0.05fs) requires 28 matrix-vector 
%mulplicationi(H*x) on average. Compared with S-RK4 simulation, which uses
%4 H*x to simulate 0.5as, PTCN is 14.2x faster. 

\begin{figure}[h]
  \begin{center}
    \subfloat[BOMD and RT-TDDFT energy transfer with different initial
    kinetic
    energies.]{\includegraphics[angle=270,width=0.50\textwidth]{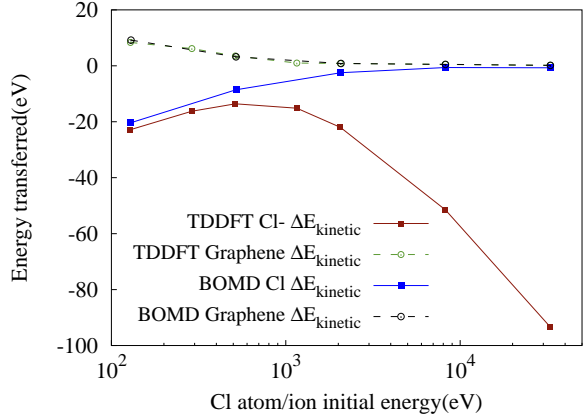}}\\
    \subfloat[Density of state after the ion collision. Green dashed
    line: Fermi energy.]{\includegraphics[angle=270,width=0.50\textwidth]{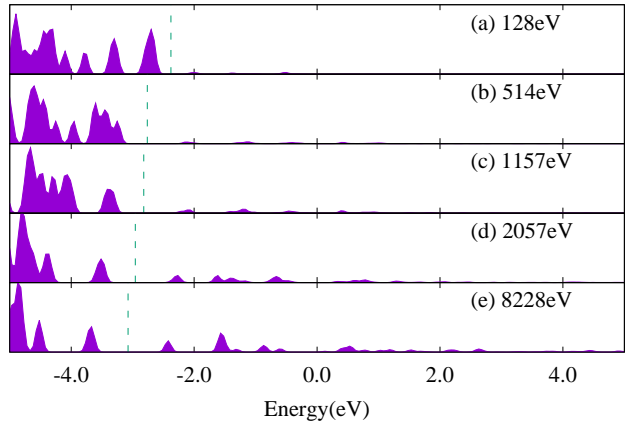}} \\
  \end{center}
  \caption{Energy transfer and density of states.
%  \LL{(a) Add BOMD result. (b) Total electronic energy should be
%  Potential energy. Cl- ion/atom should be Cl-ion / atom kinetic energy
%  for consistency. Why red color for y-axis? Change fonts to be consistent 
%  in terms of type and size.}
%  \WL{(a) added.
%      (b) Font and legend label changed.  the red color is corresponding 
%          to the red line, which is the kinetic energy of Cl atom. 
%          Potential energy are labeled green on the figure. 
  }
  \label{fig:energy}
\end{figure}

We compare the result obtained from the Ehrenfest dynamics with that
from the Born-Openheimer Molecular Dynamics (BOMD). In the BOMD
simulation, since the extra electron of Cl$^{-}$ will localize on the
conduction band of the graphene conduction rather than on Cl during the
self-consistent field iteration,  we replace the Cl$^{-}$ ion by the Cl
atom.   
%\LL{what is the criterion for computing the kinetic
%energy of the graphene? Does it depend on
%the termination time?} \WL{ Yes, the energy taken by the graphene
%depends on the time, but it deos not fluctualtes much. The 
%stoppting criteria is that we simulate the same time for both 
%TD and BOMD. Graphene $\Delta$Ek= $\Delta Ek(Cl)$ - $\Delta E_{pot}$, where
%$Delta$ E = $E_{init}$ - $E_{finalstep}$ }
% Despite the difference in the setup, we find
% that as the initial kinetic energy increases and is below 500 eV,
% both TDDFT and BOMD predict that the loss of the kinetic energy of Cl$^{-}$ (Cl atom)
% should decrease as in Fig.~\ref{fig:energy} (a). This is because most of the electrons stay in the
% ground state of the graphene nanoflake, and the kinetic energy of the
% Cl$^{-}$ (Cl atom) is mostly transferred to the kinetic energy of the
% carbon atoms during the collision process. Since the increase of the
% initial kinetic energy reduces the collision time, the energy transfer
% therefore decreases as well. This is confirmed from the decreases of the
% gain of the kinetic energy of graphene in Fig.~\ref{fig:energy} (a). On
% the other hand, when the initial kinetic energy is above $1000$ eV, BOMD
% predicts that the energy transfer continues to decrease, while TDDFT
% predicts that the energy transfer should significantly increase instead.
Fig.~\ref{fig:energy} (a) illustrates the energy transfer with different 
initial kinetic energies. 
As the Cl/Cl$^-$ initial kinetic energy increases, the gain of the kinetic energy 
by the graphene atoms decreases due to that Cl/Cl$^-$ can pass through the system faster. 
When the initial kinetic energy of Cl/Cl$^-$  is smaller than 500 eV, 
the losses of the kinetic energy for Cl/Cl$^-$ are similar between RT-TDDFT and BOMD. 
However, when the initial kinetic energy of Cl/Cl$^-$  further increases, the RT-TDDFT 
predicts an increase of the loss of the Cl/Cl$^-$ kinetic energy, while the gain of 
the graphene kinetic energy remains decreasing. 
%\DA{The remaining part of this paragraph is of a different order with 
%Lin-Wang's comment. I think this way may be clearer.}
This is a consequence of the electron excitation, which is absent in the 
BOMD simulation. 
Such excitation is illustrated in Fig.~\ref{fig:energy} (b) 
for the occupied electron density of states in the higher energy regimes. 
The occupied density of states is calculated as 
$\rho(\varepsilon):=\sum_{j=1}^{N_{e}} \sum_{i=1}^{\infty}
|\langle\phi_i(T) | \psi_j(T)\rangle|^2 \tilde{\delta}(\varepsilon-\varepsilon_{i}(T))$.  Here $\psi_{j}(T)$
is the $j$-th orbital obtained at the end of the RT-TDDFT simulation at
time $T$, and $\varepsilon_{i}(T),\phi_i(T)$ are
the eigenvalues and wavefunctions corresponding to the Hamiltonian at
time $T$. $\tilde{\delta}$ is a Dirac-$\delta$ function with a Gaussian
broadening of $0.05$ eV.

%Fig.~\ref{fig:energy} (a) shows that the loss of the kinetic energy of
%Cl$^{-}$

%kinetic energy lost by the Cl ion(atom) and the kinetic energy taken by the 
%graphene. In BO-MD simulation, the energy transferred from into all carbon
%atoms decreases with increasing projectile energy. This trend, however, does
%not apply for the total energy transferred to the target, which is defined 
%by the difference between Cl- ion initial and final kinetic energy. On the 
%other hand, the kinetic energy transfer predicted by the TDDFT and BO-MD are
%nearly indistinguishable. TDDFT simulation indicates a new trend, namely, 
%an increase of the energy transfer above 1keV. This trend reveral, also 
%observed in experiment, cannot be explained by BO-MD simulation where 
%electronic excitation is not evaluated. 

\begin{figure}[h]
  \begin{center}
    {\includegraphics[angle=270,width=0.60\textwidth]{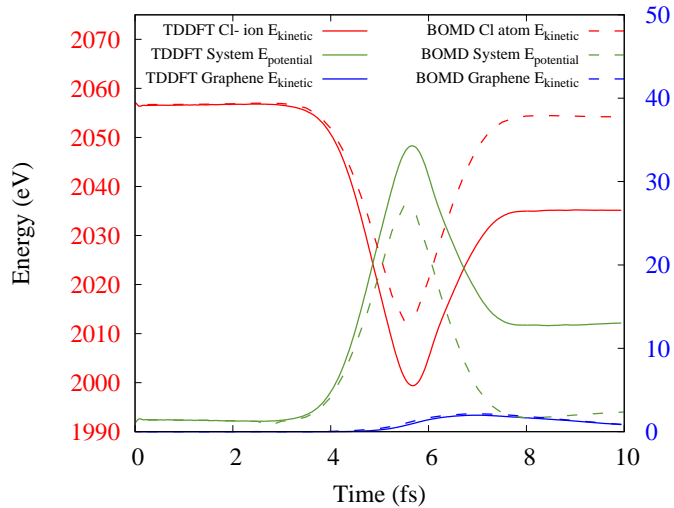}}
  \end{center}
  \caption{BOMD and RT-TDDFT energy transfer with time, projectile speed is 2.0 Bohr/fs.}
  \label{fig:energydetail}
\end{figure}

Fig.~\ref{fig:energydetail} presents further details of the energy
transfer along the trajectory of the RT-TDDFT and BOMD simulation when the
initial velocity is  2.0 Bohr/fs (2057 eV).  
%The total electronic energy of the system, Cl ion kinetic energy and
%nuclear kinetic energy of the substrate are shown with time. Total
%energy is the sum of the three energies, which is a straight line(not
%shown in the figure). 
%During the collision, the loss of the kinetic energy for the Cl
%atom(ion) is $44\sim 58$ eV kinetic energy.
%\LL{where does this number come from?}\WL{ BOMD lost 44eV in the collision
%and TDDFT lost 58eV during the collision. In the end, the 44 eV is given
%back to Cl in BOMD, but not in TDDFT. This sentence and the next are trying
%to state the same facts. we can remove this one.}
When the collision occurs at around 
$T=6$ fs, the loss of the Cl/Cl$^{-}$ kinetic energy is 44 eV and 58 eV 
under RT-TDDFT and BOMD, respectively. However, after collision Cl regains
almost all the kinetic energy in BOMD, and the final kinetic energy is only 2.5
eV less than the initial one. Correspondingly,
the kinetic energy of the graphene increases by $0.86$ eV and the
potential energy increases by 1.63 eV.  On the other hand, RT-TDDFT predicts
that the Cl$^{-}$ ion should lose $22.5$ eV kinetic energy, which is
mostly transferred to the potential energy of the excited electrons.
%\DA{at the excited orbitals 
%after the collision (Here Lin-Wang made a comment, but it seems outside 
%the scanning area and I fail to read it. 
%The modification I made here is just my guess)} 
The increase of the kinetic
energy of the graphene is $0.84$ eV and is similar to the BOMD result. 
Therefore, in RT-TDDFT, the Cl$^{-}$ loses its kinetic energy to electron excitation 
in graphene. 
% Therefore as the kinetic energy of the Cl$^{-}$ ion increases, the
% mechanism for the energy transfer changes qualitatively. It is dominated
% by the electron excitation process, and the energy transfer is mainly
% between the kinetic energy and the electronic potential energy.

%only  energy is converted to the 
%kinetic energy of the Graphene and others are given back to
%the Cl atom at the end of the simulation. In TDDFT simulation, however, 
%the Cl- ion lost $22.5$eV kinetic energy and only $0.84$eV is transferred
%to the kinetic energy of the substrate in the end. This is caused
%by the excitation of the electrons, which cannot be captured in the 
%BO-MD simulation. 
%
%One thing we observe is that, the transfer electronic energy is 
%propotional to the velocity of the Cl ion between $1-30$keV. 
% 

%\begin{figure} \label{fig:energy}
%\begin{subfigure}{.5\textwidth}
%\includegraphics[angle=270,width=0.7\textwidth]{energy_trans.eps}
%\label{fig:etrans}
%\end{subfigure}
%
%\begin{subfigure}{.5\textwidth}
%\includegraphics[angle=270,width=0.45\textwidth]{td_bo_energy.eps}
%\label{fig:td_bo_energy}
%\end{subfigure}
%
%\caption{The kinetic energy of the Cl- atom traveling through 
%the 112 atom Graphene substrate at a initial speed of $1.0 Bohr/fs$.}
%\end{figure}

\section{Conclusion}

In this work, we demonstrate that one significant factor leading to the
very small time step size in RT-TDDFT calculations is the non-optimal gauge
choice in the Schr\"odinger dynamics. Since all physical observables
should be gauge-independent, we may optimize the gauge choice to improve
the numerical efficiency without sacrificing accuracy. 
%We find that the
%optimal gauge is given by a parallel transport formalism, and the
%variation of the orbitals in this parallel transport gauge can vary
%substantially slower than that in the Schr\"odinger gauge, 
%which can be beneficial to any RT-TDDFT integrator. 
The resulting scheme can be beneficial to any RT-TDDFT integrator, and
can even be nearly symplectic. 
%For implicit time integrators, the resulting scheme can employ a much
%large time step even with a complete basis set such as planewaves. 
With the increased time step size, we hope that RT-TDDFT can be used to study 
many ultrafast problems unamenable today. 
% With the increased time step size, we hope that RT-TDDFT will soon become
% part of the routine toolkit for studying excited state properties of
% electrons for a wide range of chemical and material systems, as
% \textit{ab initio} molecular dynamics has achieved for studying ground state
% properties.

% In this work, we demonstrate that one significant factor leading to the
% very small time step in RT-TDDFT calculations is the non-optimal gauge
% choice in the Schr\"odinger dynamics. Since all physical observables
% should be gauge-independent, we may optimize the gauge choice to improve
% the numerical efficiency without sacrificing accuracy. We find that the
% optimal gauge is given by a parallel transport formalism, and the
% variation of the orbitals in this parallel transport gauge can vary
% substantially slower than that in the Schr\"odinger gauge. When combined
% with implicit time integrators, the resulting scheme can employ a much
% large time step even with a complete basis set such as planewaves. With
% the increased time step size, we expect that RT-TDDFT will soon become
% part of the routine toolkit for studying excited state properties of
% electrons for a wide range of chemical and material systems, as
% \textit{ab initio} molecular dynamics has achieved for studying ground state
% properties.

\section*{Acknowledgement}

This work was partially supported by the National Science Foundation
under Grant No. 1450372, No. DMS-1652330  (D. A., W. J. and L. L.), and
by the Department of Energy under Grant No. DE-SC0017867, No.
DE-AC02-05CH11231 (L. L.), and by the Department of Energy Theory of
Materials (KC2301) program under Contract No. DE-AC02-05CH11231 (L. W.). 
We thank the National Energy Research
Scientific Computing (NERSC) center and the Berkeley Research Computing
(BRC) program at the University of California, Berkeley for making
computational resources available.  We thank Stefano Baroni, Roberto
Car, Zhanghui Chen, Wei Hu, Christian Lubich and Chao Yang for helpful
discussions.

\appendix

\section{Appendix A: Derivation of the parallel transport gauge} 

%The parallel transport gauge is the solution to the following variational problem
%\begin{equation}\label{eqn:minproblem}
%  \min_{U(t)} \quad \norm{\dot{\Phi}}^2_{F},
%  \ \text{s.t.} \ \Phi(t) = \Psi(t)U(t), U^{*}(t)U(t)=I_{N_{e}}.
%\begin{split}
%  \min_{U(t)} \quad &\norm{\dot{\Phi}}^2_{F} \\
%  \text{s.t.} \quad &\Phi(t) = \Psi(t)U(t), \quad U^{*}(t)U(t)=I.
%\end{split}
%\end{equation}
%Here $\|\dot{\Phi}\|_F^2 := \text{Tr}[\dot{\Phi}^*\dot{\Phi}]$ measures the Frobenius 
%norm if the time derivative of the transformed orbitals. 
In order to solve~\eqref{eqn:minproblem}, we first split $\dot{\Phi}$ into two 
orthogonal components 
\begin{equation}
    \dot{\Phi} = P\dot{\Phi} + (I-P)\dot{\Phi} {.}
\end{equation}
Then we have
\begin{equation}\label{eqn:orthogonal}
    \|\dot{\Phi}\|_F^2 = \|P\dot{\Phi}\|_F^2 + \|(I-P)\dot{\Phi}\|_F^2 {.}
\end{equation}
To reformulate the second term, we take the time derivative on the equation $P\Phi = \Phi$ and get
\begin{equation}
    \dot{P}\Phi = \dot{\Phi} - P\dot{\Phi} = (I-P)\dot{\Phi} {.}
\end{equation}
Thus Eq.~\eqref{eqn:orthogonal} becomes
\begin{equation}\label{eqn:orthogonal2}
    \|\dot{\Phi}\|_F^2 = \|P\dot{\Phi}\|_F^2 + \|\dot{P}\Phi\|_F^2
     = \|P\dot{\Phi}\|_F^2 + \|\dot{P}\Psi\|_F^2 {,}
\end{equation}
where the last equality comes from that $\Phi = \Psi U$ 
and $U$ is a unitary gauge matrix. 

Eq.~\eqref{eqn:orthogonal2} has a clear physical interpretation. The second 
term 
\begin{equation}
\|\dot{P}\Psi\|_F^2 = \text{Tr}[\Psi^*\dot{P}^2\Psi] 
= \text{Tr}[\dot{P}^2\Psi\Psi^*] = \text{Tr}[\dot{P}^2P]
\end{equation}
is defined solely from the density matrix and is thus gauge-invariant. 
Therefore the variation of $\Phi$ is minimized when 
\begin{equation}\label{eqn:PTcondition_2}
    P\dot{\Phi} = 0 {,}
\end{equation}
which is exactly the parallel transport condition. 

Now we would like to directly write down the governing equation of $\Phi$. 
First, the equation $\Phi = P\Phi$ and the parallel transport condition~\eqref{eqn:PTcondition} imply that 
\begin{equation}
    \dot{\Phi} = \partial_t(P\Phi) = \dot{P}\Phi + P\dot{\Phi} = \dot{P}\Phi {.}
\end{equation}
Together with the von Neumann equation, we have 
\begin{equation}\label{eqn:pt_2}
    \begin{split}
        \I \dot{\Phi} &= \I \dot{P}\Phi = [H,P]\Phi 
        = HP\Phi - PH\Phi = H\Phi - \Phi(\Phi^*H\Phi) {.}
    \end{split}
\end{equation}
This is exactly the parallel transport dynamics. 

The name ``parallel transport gauge'' originates from the parallel transport
 formulation associated with a family of density matrices $P(t)$, which
generates a parallel transport evolution operator $\mathcal{T}(t)$ as (see \eg ~\cite{Nakahara2003,CorneanMonacoTeufel2017})
 \begin{equation}
   \I \partial_t \mathcal{T} = [\I \partial_{t} P,P] \mathcal{T}, \quad \mathcal{T}(0)=I.
   \label{eqn:ptevolve}
 \end{equation}
 We demonstrate that starting from an initial set of orbitals $\Psi_{0}$, the solution
 to the parallel transport dynamics~\eqref{eqn:pt} is simply evolved by 
 the parallel transport evolution operator according to 
 $\Phi(t)=\mathcal{T}(t)\Psi_{0}$. 
 To show this, we first prove the following relation
 \begin{equation}\label{eqn:ptcommute}
     P(t)\mathcal{T}(t) = \mathcal{T}(t)P(0)
 \end{equation}
by showing that both sides solve the same initial value problem. 
Note that $\mathcal{T}(t)P(0)$ satisfies 
\begin{equation}
    \I \partial_t(\mathcal{T}(t)P(0)) = [\I \partial_{t} P,P] (\mathcal{T}(t)P(0)){.}
\end{equation}
We then would like to derive the differential equation $P(t)\mathcal{T}(t)$ satisfies. 
Taking the time derivative on both sides of the identity $P = P^2$, we have 
  \begin{equation}
    \dot{P} = \dot{P} P + P \dot{P}
    \label{eqn:Pidentity}
  \end{equation}
  and thus
  \begin{equation}
    P\dot{P}P = (\dot{P}-\dot{P}P)P = \dot{P}(P-P^2) = 0{.}
  \end{equation}
  Then
  \[
  \I \partial_t(P\mathcal{T}) = \I \dot{P}\mathcal{T} + 
  \I P\dot{\mathcal{T}} = \I \dot{P}\mathcal{T} 
  + \I P[\dot{P},P]\mathcal{T} = \I \dot{P} P\mathcal{T}.
  \]
  On the other hand,
  \[
  [i\dot{P},P](P\mathcal{T}) = \I (\dot{P} PP\mathcal{T} -
  P\dot{P}P\mathcal{T}) = \I \dot{P} P\mathcal{T}.
  \]
  Therefore
  \begin{equation}
    \I \partial_t (P\mathcal{T}) = [i\dot{P},P] (P\mathcal{T}).
    \label{eqn:PTopediff}
  \end{equation}
  Together with the same initial value $P(0)\mathcal{T}(0)=\mathcal{T}(0)P(0)=P(0)$, 
  we have proved $P(t)\mathcal{T}(t) = \mathcal{T}(t)P(0)$. 
  Using this relation, we have
  \begin{equation}
    P(t)(\mathcal{T}(t)\Psi_0) = \mathcal{T}(t)P(0)\Psi_0 = 
    \mathcal{T}(t)\Psi_0.
    \label{eqn:Pphi1}
  \end{equation}
  Since $\mathcal{T}(t)$ is unitary, we have
  $(\mathcal{T}(t)\Psi_0)^*(\mathcal{T}(t)\Psi_0) = I$ for all $t$. Hence 
  $\mathcal{T}(t)\Psi_0$ forms an orthogonal basis in the image of $P(t)$. 
  Therefore 
  \begin{equation}
    P(t) = (\mathcal{T}(t)\Psi_0)(\mathcal{T}(t)\Psi_0)^*.
    \label{eqn:Pphi2}
  \end{equation}
  By Eq.~\eqref{eqn:ptcommute},~\eqref{eqn:PTopediff} 
  and the von Neumann equation, we have
  \begin{equation}
    \begin{split}
    \I \partial_{t}(\mathcal{T}\Psi_0) =& \I \partial_{t}(P\mathcal{T}) \Psi_0
    = [\I \dot{P},P] P \mathcal{T} \Psi_0 \\
    = &\I \dot{P}P\mathcal{T} \Psi_0
    = HP\mathcal{T} \Psi_0 - PHP\mathcal{T} \Psi_0.
    \end{split}
    \label{}
  \end{equation}
  Finally using
  Eq.~\eqref{eqn:Pphi1} and~\eqref{eqn:Pphi2}, we have
  \[
  \I \partial_{t}(\mathcal{T}\Psi_0) = H(\mathcal{T} \Psi_0) - 
  (\mathcal{T} \Psi_0)((\mathcal{T} \Psi_0)^*H(\mathcal{T} \Psi_0)) {,}
  \]
  thus $\mathcal{T} \Psi_0$ precisely solves the parallel transport dynamics, 
  indicating $\Phi(t) = \mathcal{T}(t) \Psi_0$.

\section{Appendix B: Time discretization schemes}

We list several propagation schemes used in this paper, but the parallel
transport dynamics can be discretized with any propagator.
Here all the $H_n = H(t_n,P_n)$ is the Hamiltonian at step $t_n$, 
and $t_{n+\frac{1}{2}} = t_n + \frac{1}{2}\Delta t$, $t_{n+1} = t_n + \Delta t$. 
For implicit time integrators, $\Psi_{n+1}$ or $\Phi_{n+1}$ needs to be solved 
self-consistently. 

The standard explicit 4th order Runge-Kutta scheme for the Schr\"odinger dynamics (S-RK4): 
\begin{equation}
    \begin{split}
        k_1 &= -\I \Delta t H_n\Psi_n, \\
        \Psi_n^{(1)} &= \Psi_n + \frac{1}{2}k_1, 
        \quad H_n^{(1)} = H(t_{n+\frac{1}{2}},\Psi_n^{(1)}\Psi_n^{(1)*})\\ 
        k_2 &= -\I \Delta t H_n^{(1)}\Psi_n^{(1)}, \\
        \Psi_n^{(2)} &= \Psi_n + \frac{1}{2}k_2,
        \quad H_n^{(2)} = H(t_{n+\frac{1}{2}},\Psi_n^{(2)}\Psi_n^{(2)*})\\
        k_3 &= -\I \Delta t H_n^{(2)}\Psi_n^{(2)}, \\
        \Psi_n^{(3)} &= \Psi_n + k_3, 
        \quad H_n^{(3)} = H(t_{n+1},\Psi_n^{(3)}\Psi_n^{(3)*})\\
        k_4 &= -\I \Delta t H_n^{(3)}\Psi_n^{(3)}, \\
        \Psi_{n+1} &= \Psi_n + \frac{1}{6}(k_1 + 2k_2 + 2k_3 + k_4).
    \end{split}
\end{equation}

The standard explicit 4th order Runge-Kutta scheme for the parallel transport dynamics (PT-RK4): 
\begin{equation}
    \begin{split}
        k_1 &= -\I \Delta t \{H_n\Phi_n-\Phi_n(\Phi_n^*H_n\Phi_n)\}, \\
        \Phi_n^{(1)} &= \Phi_n + \frac{1}{2}k_1, 
        \quad H_n^{(1)} = H(t_{n+\frac{1}{2}},\Phi_n^{(1)}\Phi_n^{(1)*})\\ 
        k_2 &= -\I \Delta t \{H_n^{(1)}\Phi_n^{(1)} - \Phi_n^{(1)}(\Phi_n^{(1)*}H_n^{(1)}\Phi_n^{(1)})\}, \\
        \Phi_n^{(2)} &= \Phi_n + \frac{1}{2}k_2,
        \quad H_n^{(2)} = H(t_{n+\frac{1}{2}},\Phi_n^{(2)}\Phi_n^{(2)*})\\
        k_3 &= -\I \Delta t \{H_n^{(2)}\Phi_n^{(2)} - \Phi_n^{(2)}(\Phi_n^{(2)*}H_n^{(2)}\Phi_n^{(2)})\}, \\
        \Phi_n^{(3)} &= \Phi_n + k_3, 
        \quad H_n^{(3)} = H(t_{n+1},\Phi_n^{(3)}\Phi_n^{(3)*})\\
        k_4 &= -\I \Delta t \{H_n^{(3)}\Phi_n^{(3)} - \Phi_n^{(3)}(\Phi_n^{(3)*}H_n^{(3)}\Phi_n^{(3)})\}, \\
        \Phi_{n+1} &= \Phi_n + \frac{1}{6}(k_1 + 2k_2 + 2k_3 + k_4).
    \end{split}
\end{equation}

The implicit Crank-Nicolson scheme for the Schr\"odinger dynamics (S-CN):
\begin{align}\label{eqn:scn}
    \left(I + \I \frac{\Delta t}{2} H_{n+1}\right)\Psi_{n+1}
    = \left(I -\I \frac{\Delta t}{2} H_n\right)\Psi_n.
\end{align}

The implicit Crank-Nicolson scheme for the parallel transport dynamics (PT-CN):
\begin{equation}\label{eqn:ptcnApp}
  \begin{split}
    &\Phi_{n+1} + \I \frac{\Delta t}{2} \left\{
    H_{n+1} 
    \Phi_{n+1} - \Phi_{n+1}\left(\Phi_{n+1}^{*}
    H_{n+1} \Phi_{n+1}\right)\right\} \\
    = &\Phi_{n} -\I \frac{\Delta t}{2} \left\{
    H_{n}
    \Phi_{n} - \Phi_{n}\left(\Phi_{n}^{*}
    H_{n} \Phi_{n}\right)\right\}.
  \end{split}
\end{equation}

\section{Appendix C: Details of RT-TDDFT calculations}

\begin{figure}[h]
\begin{center}
\includegraphics[width=0.4\textwidth]{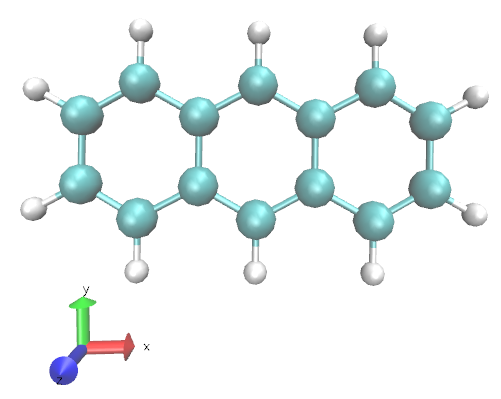}
\end{center}
\caption{Atomic configuration of anthracene.}
\label{fig:anthracene}
\end{figure}

For the example of absorption spectrum of anthracene (C$_{14}$H$_{10}$,
Fig.~\ref{fig:anthracene}).
The simulation is performed using a cubic supercell of size
$(20\angstrom)^3$, and the kinetic energy cutoff is $20$ au. 
In order to compute the absorption spectrum, a
$\delta$-pulse of strength $0.005$ au is applied to the $x,y,z$
directions
to the ground state wavefunctions respectively, and the system
is then propagated for $4.8$ fs along each direction. This gives the
polarization tensor $\chi(\omega)$, and the optical absorption cross-section
is evaluated as 
$$\sigma(\omega) = (4\pi \omega/c) \Im \Tr[\chi(\omega)].$$

\begin{figure}[h]
  \begin{center}
    \subfloat[]{\includegraphics[width=0.45\textwidth]{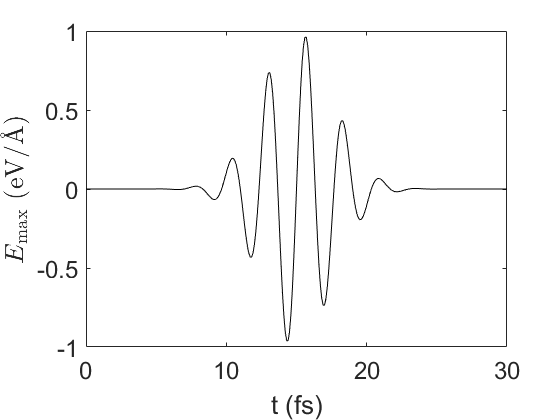}}
    \subfloat[]{\includegraphics[width=0.45\textwidth]{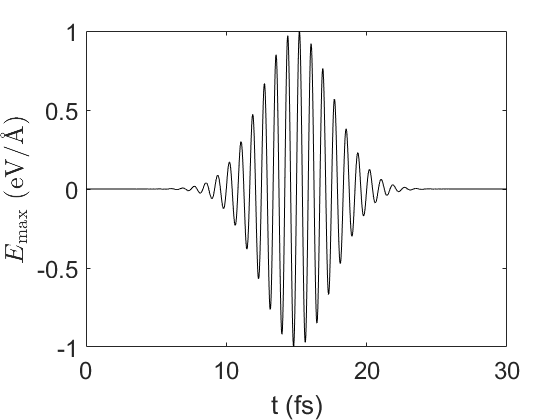}}
  \end{center}
  \caption{External fields of the lasers. 
  The peak electric field is 1.0 eV/\r A,
occurring at $t_0 = 15.0$ fs, and the FWHM pulse width is 6.0 fs. 
The wavelength is 800 nm in (a), 250 nm in (b). }
  \label{fig:benzeneVext}
\end{figure}

For the example of the benzene molecule driven by an ultrashort laser
pulse, the electric field takes the form
\begin{equation}
\textbf{E}(t) = \hat{\textbf{k}}E_{\text{max}}\exp \Big[-\frac{(t-t_0)^2}{2a^2}\Big]\sin [\omega(t-t_0)] {,}
\end{equation}
where $\hat{\textbf{k}}$ is a unit vector defining the polarization of the electric field. 
The parameters $a,t_0,E_{\text{max}},\omega$ define
the width, the initial position of the center,
the maximum amplitude of the Gaussian envelope,
and the frequency of the laser, respectively.  In practice $\omega$ and
$a$ are often determined by
the wavelength $\lambda$ and the full width at half maximum (FWHM) pulse width~\cite{RussakoffLiHeVarga2016}, 
\ie $\lambda \omega = 2\pi c$ and $\text{FWHM} = 2a\sqrt{2\log 2}$,
where $c$ is the speed of the light. 
In this example, the peak electric field $E_{\text{max}}$ is 1.0 eV/\r A,
occurring at $t_0 = 15.0$ fs. 
The FWHM pulse width is 6.0 fs,
and the polarization of the laser field is aligned along the $x$ axis 
(the benzene molecule is in $x$-$y$ plane). 
We consider one relatively slow laser with wavelength 800 nm, and another faster
laser with wavelength 250 nm, respectively
(Fig.~\ref{fig:benzeneVext}). 
The electron dynamics
for the first laser is in the near adiabatic regime, where the system
stays near the ground state after the active time interval of the laser,
while the second laser drives a significant amount of electrons to
excited states. We propagate RT-TDDFT to $T = 30.0$ fs.  
For the parameters in the Anderson mixing, the step length $\alpha$ is $0.2$,
the mixing dimension is 10, and the tolerance is $10^{-6}$.

Even though the electron dynamics varies rapidly under the
laser with 250 nm wavelength, PT-CN can still be stable within a
relatively large range of time steps. 
Table~\ref{tab:Example_TDDFT_EnergyDifference} measures the accuracy of
PT-CN with $\Delta t=$ 5 as, 6.5 as, 7.5 as, 10 as and 20 as, respectively. 
We find that the number of matrix-vector multiplications per orbital systematically reduces
as the step size increases. When $\Delta t=$ 20 as, the speed up
over S-RK4 is $12.6$, and this is at the expense of overestimating
the energy by $0.0672$ eV after the active interval of the laser.  
Hence one may adjust the time step
size to obtain a good compromise between efficiency and accuracy, 
and use PT-CN to quickly study the electron dynamics with a
large time step. This is not possible using an explicit scheme
like S-RK4.

\begin{table}
  \centering
  \begin{tabular}{cc|cc|cc}
  \hline Method & $\Delta t$ (fs) & AEI (eV) & AOE (eV) & MVM & Speedup\\\hline
  S-RK4 & 0.0005 & 0.5260  & /       & 152000 & /    \\
  PT-CN & 0.005  & 0.5340  & 0.0080  & 28610  & 5.3  \\
  PT-CN & 0.0065 & 0.5347  & 0.0087  & 22649  & 6.7  \\
  PT-CN & 0.0075 & 0.5362  & 0.0102  & 21943  & 6.9  \\
  PT-CN & 0.01   & 0.5435  & 0.0175  & 15817  & 9.6  \\
  PT-CN & 0.02   & 0.5932  & 0.0672  & 12110  & 12.6 \\\hline
  \end{tabular}
  \caption{Accuracy and efficiency of PT-CN for the electron dynamics
  with the 250 nm laser compared to S-RK4.
  The accuracy is measured using 
  the average energy increase (AEI) after 25.0 fs and
  the average overestimated energy (AOE) after 25.0 fs.
  The efficiency is measured using 
  the total number of matrix-vector multiplications per orbital (MVM) 
  during the time interval from 5.5 fs to 24.5 fs, 
  and the computational speedup.
  }\label{tab:Example_TDDFT_EnergyDifference}
\end{table}

%For the ion collision example, the system is shown in
%Fig.~\ref{fig:grnmodel}\LL{two pictures, one for the overall atomic
%configuration and the current one}. A 113 atom(including the Cl-) 7x4 superlattice
%is built with  periodic boundary condition and sampling with $\Gamma$
%point only. The Z direction is 12 \angstrom. An energy cutoff of 30
%au is used. The system has 228 states and each state is occupied
%with 2 electrons. \LL{Discuss the setup of the initial electronic state
%of the system with the Cl ion. The termination time for TDDFT
%calculations with different initial velocity. }
%
%\begin{figure}
%\begin{center}
%\includegraphics[width=0.25\textwidth]{112GRN_x.png}
%\end{center}
%\caption{ Model of Cl-Graphene collision. Red point is the center of the 
%graphene ring.}
%\label{fig:grnmodel}
%\end{figure}

\begin{figure}[h]
  \begin{center}
    \subfloat[$x$ direction]{\includegraphics[width=0.45\textwidth]{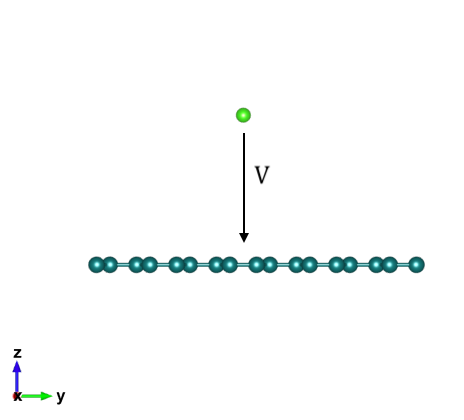}}
    \subfloat[$z$ direction]{\includegraphics[width=0.45\textwidth]{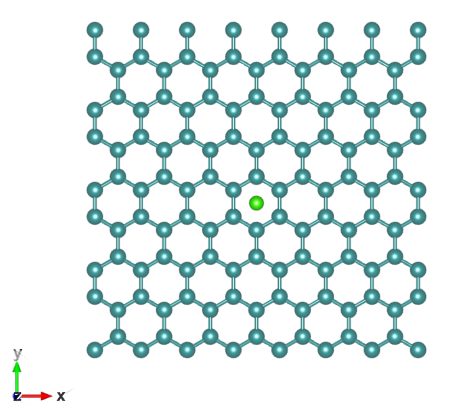}}
  \end{center}
  \caption{Model for the collision of Cl/Cl$^{-}$ and a graphene
  nanoflake.
%  Cl$^{-}$ ion(in green) is traveling through the hexagonal ring center.
   }
  \label{fig:grnmodel}
\end{figure}

For the ion collision example, the system is shown in
Fig.~\ref{fig:grnmodel}. The supercell has 113 atoms (including Cl$^{-}$)
%consisting of
%$7\times 4$ unit cells of graphene
%is built with periodic boundary condition 
sampled at the $\Gamma$ point. 
The length of the box along the $z$ direction is 12 \angstrom and the Cl$^{-}$ ion is placed
6 \angstrom away from the graphene nanoflake. The initial velocity is
perpendicular to the
center of one hexagonal ring of the graphene nanoflake. The kinetic energy cutoff is
30 Hartree.
The system has 228 states and each state is occupied with 2 electrons,
and the calculation is performed with 228 computational cores at NERSC. 
%In order to obtain the correct initial electronic structure for
%Cl$^{-}$, we first perform compute
%two separated systems of Cl$^{-}$ and graphene to obtain the converged wave functions.
%The initial setup of the TDDFT wave functions is to combine the Cl$^{-}$ 
%and graphene wavefunctions by aligning them.
%Then the Cl- wave function is orthogonalized to the graphene wave function to 
%get the intitial wave functions of the TDDFT system. 
%The initial setup is created by combining two converged systems of
%ground state Cl- and graphene wave functions. 
The initial velocity of Cl$^{-}$ ion is set to be $0.5, 0.75, 1.0, 2.0, 4.0, 8.0$ Bohr/fs. 
We stop the simulation after Cl$^{-}$ passes through the graphene sheet,
so the total simulation 
time is $40, 26.7, 20, 10, 5, 2.5$ fs, respectively. The time step size for PT-CN method 
is set to be $50, 50, 50, 25, 12.5, 6.25$ as in the simulation
accordingly. 
%The stopping criteria is set to $1.0E-5$ in all simulation. 
%It takes $3.5$ to $22$ hours to finish the TDDFT simulation 
%for different initial velocity. 

We find that PT-CN method is again more efficient compared to the S-RK4 method. For example, 
in the case of $1.0$ Bohr/fs test, the time step for PT-CN can be chosen to be $50$ as, 
while the time step for the S-RK4 method cannot be chosen to be larger
than $0.5$ as. 
PT-CN needs 27 matrix-vector multiplications per orbital for each RT-TDDFT
step, and S-RK4 needs 4. So PT-CN is around $14$ times faster than S-RK4 for this 
test.

% \begin{table}
%   \centering
%   \begin{tabular}{cc|cc|cc}
%   \hline Vel & STime & $\Delta t$ & MVM & CT (hour) & MVM \\\hline
%   0.5  & 40.0  & 0.05    & 0.0080  & 21.9  & 29  \\
%   0.75 & 26.7  & 0.05    & 0.0087  & 15    & 27  \\
%   1.0  & 20.0  & 0.05    & 0.0102  & 11    & 30  \\
%   2.0  & 10.0  & 0.025   & 0.0175  & 4.7   & 28  \\
%   4.0  & 5.0   & 0.0125  & 0.0175  & 4.0   &   \\
%   8.0  & 2.5   & 0.00625 & 0.0672  & 3.4   & 12.6 \\\hline
%   \end{tabular}
%   \caption{Accuracy and efficiency of PT-CN for the electron dynamics
%   with the 250 nm laser compared to S-RK4.
%   The accuracy is measured using 
%   the average energy increase (AEI) after 25.0 fs and
%   the average overestimated energy (AOE) after 25.0 fs.
%   The efficiency is measured using 
%   the total number of matrix-vector multiplications (MVM) 
%   during the time interval from 5.5 fs to 24.5 fs, 
%   and the computational speedup.
%   }\label{tab:Example_TDDFT_EnergyDifference}
% \end{table}

%\LL{Discuss the setup of the initial electronic state
%of the system with the Cl ion. The termination time for TDDFT
%calculations with different initial velocity. }

%\begin{figure}
%\begin{center}
%\includegraphics[width=0.25\textwidth]{112GRN.png}
%\end{center}
%\caption{ Model of Cl-Graphene collision. Red point is the center of the
%graphene ring.}
%\label{fig:grnmodel}
%\end{figure}

\bibliography{ptref}
\newpage

\end{document}